\documentclass[twocolumn, prl, aps, floats]{revtex4}

\usepackage{amsmath, amssymb, bbm, bm, calc, color, dblfloatfix, epsf, epsfig, float, graphicx, mathtools, nicefrac, ragged2e, relsize, setspace, tabularx, yfonts, xcolor}
\def\f{F}
\def\A{\mathbf{A}}

\def\v{\mathbf{v}}
\def\vi{\mathbf{v_0}}
\def\vii{\mathbf{v_1}}

\def\p{\mathbf{p}}
\def\q{\mathbf{q}}

\def\na{\overline{a}}
\def\nb{\overline{b}}

\begin{document}

\frenchspacing
\title{Exact distribution of the output of a deep-layered machine}
\author{Thomas Fink}
\affiliation{London Institute for Mathematical Sciences, Royal Institution, 21 Albemarle St, London W1S 4BS, UK}

\begin{abstract}
\noindent
Deep learning builds complex global functions through the repeated composition of local functions. 
But the resulting distribution over global functions is unknown.
We derive this distribution exactly for finite width and depth when the local functions are chosen at random.
Surprisingly, depth first creates a sharply structured distribution, then erases this structure as probability drains into two absorbing states.
This competition produces a crossover at a depth exponential in the number of inputs, beyond which additional layers mainly drive collapse.
The transient structure may help explain why highly overparameterized networks can still generalize.
\end{abstract}

\maketitle
\begin{figure}[b!]
	\centering
	\includegraphics[width=1\columnwidth]{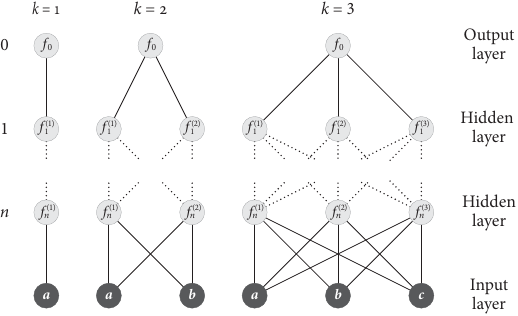}
	\caption{\footnotesize
	\textbf{Deep-layered machines.}
		A deep-layered machine takes $k$ Boolean arguments $(a,b,\ldots)$ as inputs, shown by the dark nodes. 
		Each light node applies a random Boolean function $f$ to the $k$ outputs from the layer below, 
		producing a single Boolean output. 
		Composing these functions down to depth $n$ produces the global function $F_n(a,b,\ldots)$. 
		Each $f$ is chosen independently and uniformly from the $2^{2^k}$ Boolean functions of $k$ inputs. 
		The goal of this paper is to understand the distribution of the global function $F_n$.
		\label{Architecture}
		}
\end{figure}
\begin{figure}[!b]
	\includegraphics[width=1\columnwidth]{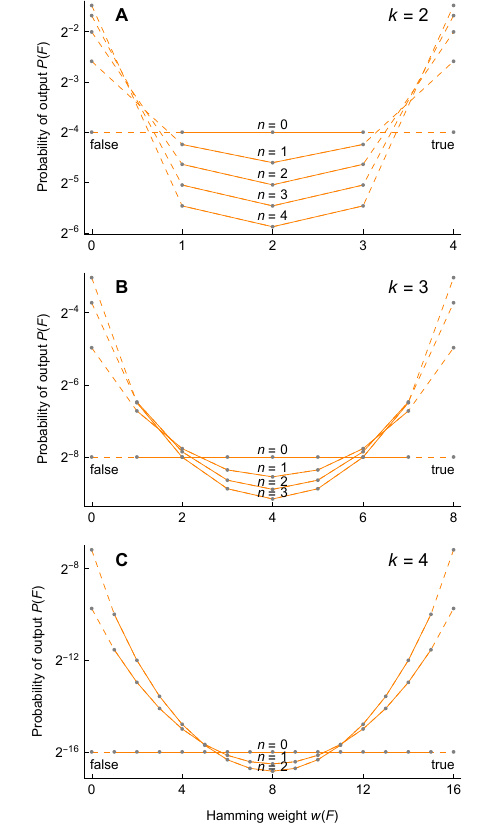}
	\caption{\small
	\textbf{Output distributions and tests of the theory.}
	We show the output function probability $P(F_i) = p_i$ against Hamming weight $w(F_i)$ 
	for small values of $k$ and $n$, including those shown in Table \ref{k123Distribution}. 
	Functions with the same Hamming weight have the same probability. 
	Starting from the uniform distribution at $n=0$, increasing depth makes the bulk of $P(F)$ {\sf U}-shaped. 
	Meanwhile, probability drains from the bulk and accumulates in the constant functions true and false.
	The curves are our exact predictions; 
	the points are independent computer experiments, 
	obtained by exhaustive enumeration where possible and Monte Carlo sampling otherwise. 
	The experiments agree exactly with theory or, when sampled, within statistical uncertainty.
	}
    \label{ComputerExperiment}
\end{figure}

\noindent
Composition is how local rules generate complex behavior, 
from biological networks to digital computation to artificial intelligence. 
In deep learning, successive layers build global functions from local operations, 
yet the resulting distribution over functions remains largely unknown.
Here we derive the exact output distribution of a deep-layered Boolean machine at finite width and depth.
The model provides a solvable setting for understanding how repeated composition reshapes function space.
\\ \indent
Random compositions of Boolean functions have been studied extensively in the context of Boolean networks \cite{Mozeika2011, Drossel2008} and the Kauffman model \cite{Samuelsson03, Drossel05a, Fink2024, Fink2023}. 
In those systems, the network architecture is random and the local Boolean functions typically depend on only a few inputs. 
Deep-layered machines represent a complementary regime: 
the architecture is highly ordered, but each node computes an arbitrary Boolean function of all of the nodes in the layer below. 
Previous work has characterized asymptotic properties of deep-layered machines \cite{Li2018, Mozeika2020}, 
in particular the collapse of the function space onto constant functions in the infinite-depth limit. 
Yet the asymptotic limit conceals the most consequential behavior:
repeated composition first creates functional structure and then destroys it.
\\ \indent
We consider a deep-layered Boolean machine with $k$ arguments and network depth $n$, in which every node is assigned a Boolean function chosen uniformly from the $2^{2^k}$ possibilities (see Fig. \ref{Architecture}). 
A key insight is that the probability of an output depends only on its Hamming weight: the number of input configurations for which the function returns true. 
This symmetry reduces the problem from a Markov process on $2^{2^k}$ states to one on just $2^k+1$ states.
We derive the exact transition matrix that takes the system from depth $n$ to $n+1$ and obtain the full spectrum of eigenvalues:
$\lambda_r = (2^k)_r/2^{kr}$.
\\ \indent
Our analysis reveals two competing effects as the depth $n$ increases. 
First, the distribution of the bulk of the output functions (everything but true and false) develops a pronounced bias: 
functions with low or high Hamming weight become exponentially more likely than those with middling weight. 
Second, probability steadily accumulates in the constant functions true and false, 
which eventually dominate the distribution. 
Crucially, these processes proceed at different rates: the bulk takes shape faster than it loses probability.
\\ \indent
Strikingly, these opposing tendencies produce a crossover at a depth of order $2^k$. 
Before this scale, added depth mainly reshapes the bulk distribution; 
beyond it, probability loss into true and false takes over, and additional layers progressively erase functional diversity. 
We confirm these predictions with independent computer experiments, 
using exhaustive enumeration for small systems and Monte Carlo sampling for larger ones.
\\ \indent
Because the model strips away learning and architectural variation, 
it isolates the effect of repeated composition on the distribution over functions. 
Even though the local functions are chosen uniformly, 
composition generates a strongly nonuniform global function distribution. 
Thus depth can shape the prior over functions before training enters.
Such function priors have recently been empirically associated with generalization in deep neural networks \cite{Mingard2025}.
\begin{table}[b!]
\begin{tabularx}{\columnwidth}{@{\extracolsep{\fill}}lcccc}
{\sf \textbf{A} ($k = 1$)}			&  \multicolumn{4}{c}{\textit{Probability of $F(a)$}}			\\ 
\hspace{2pt}\emph{Output $\f(a)$}		\hspace{10pt}  	$w$		& $n \!=\!  0$		& $n \!=\!  1$		& $n \!=\!  2$		& $n \!=\!  3$		\\ 
$ \left. \begin{array}{ccc}
	\makebox[18pt][c]{false}		\phantom{\hspace{14pt}}		& 00				& \hspace{15pt} 	0
\end{array} 	\hspace{8pt} 		\right\} $					& $\frac{1}{4}$		& $\frac{6}{4^2}$ 	& $\frac{28}{4^3}$	& $\frac{120}{4^4}$	\vspace{1pt} 	\\
$ \left. \begin{array}{ccc}
	\makebox[18pt][c]{$a$}		\phantom{\hspace{14pt}}		& 10				& \hspace{15pt} 	1 	\\
	\makebox[18pt][c]{$\na$}		\phantom{\hspace{14pt}}		& 01				& \hspace{15pt}  	1	\\
\end{array} 	\hspace{8pt} 		\right\} $					& $\frac{1}{4}$		& $\frac{2}{4^2}$ 	& $\frac{4}{4^3}$	& $\frac{8}{4^4}$	\vspace{1pt} 	\\
$ \left. \begin{array}{ccc}
	\makebox[18pt][c]{true}		\phantom{\hspace{14pt}}		& 11				& \hspace{15pt} 	2
\end{array}	\hspace{8pt} 		\right\} $					& $\frac{1}{4}$		& $\frac{6}{4^2}$	& $\frac{28}{4^3}$	& $\frac{120}{4^4}$ 	\vspace{6pt} 
\end{tabularx}
\\
\begin{tabularx}{\columnwidth}{@{\extracolsep{\fill}}lcccc}
{\sf \textbf{B} ($k = 2$)}				&  \multicolumn{4}{c}{\hspace{-11pt} \emph{Probability of $F(a, b)$}}					\\ 
\hspace{2pt}\emph{Output $\f(a,b)$}		\hspace{4pt}  	$w$	& $n \!=\!  0$		& $n \!=\!  1$		& $n \!=\!  2$		& $n \!=\!  3$		\\ 
$ \left. \begin{array}{ccc}
	\makebox[18pt][c]{false}			\phantom{\hspace{10pt}}		& 0000		& \hspace{11pt} 	0
\end{array} 	\hspace{5pt} 			\right\} $	& $\frac{1}{16}$	& $\frac{680}{16^3}$ 	& $\frac{261056}{16^5}$	& $\frac{83663360}{16^7}$	\vspace{1pt} 	\\
$ \left. \begin{array}{ccc}
	\makebox[18pt][c]{$a b$}			\phantom{\hspace{10pt}}		& 1000 		& \hspace{11pt} 	1 	\\
	\makebox[18pt][c]{$a \nb$}		\phantom{\hspace{10pt}}		& 0100		& \hspace{11pt}  	1	\\
	\makebox[18pt][c]{$\na b$}		\phantom{\hspace{10pt}}		& 0010 		& \hspace{11pt} 	1	\\
	\makebox[18pt][c]{$\na \nb$}		\phantom{\hspace{10pt}}		& 0001 		& \hspace{11pt}  	1	
\end{array} 	\hspace{5pt} 			\right\} $	& $\frac{1}{16}$	& $\frac{216}{16^3}$ 	& $\frac{42048}{16^5}$	& $\frac{8087040}{16^7}$		\vspace{1pt} 	\\
$ \left. \begin{array}{ccc}
	\makebox[18pt][c]{$a$}			\phantom{\hspace{10pt}}		& 1100 		& \hspace{11pt} 	2	\\
	\makebox[18pt][c]{$\na$}			\phantom{\hspace{10pt}}		& 0011 		& \hspace{11pt} 	2	\\
	\makebox[18pt][c]{$b$}			\phantom{\hspace{10pt}}		& 1010 		& \hspace{11pt} 	2	\\
	\makebox[18pt][c]{$\nb$}			\phantom{\hspace{10pt}}		& 0101 		& \hspace{11pt} 	2	\\
	\makebox[18pt][c]{$a \oplus b$}		\phantom{\hspace{10pt}}		& 0110 		& \hspace{11pt} 	2	\\
	\makebox[18pt][c]{$a \oplus \nb$}	\phantom{\hspace{10pt}}		& 1001 		& \hspace{11pt} 	2	
\end{array} 	\hspace{5pt} 			\right\} $	& $\frac{1}{16}$	& $\frac{168}{16^3}$		& $\frac{31680}{16^5}$	& $\frac{6068736}{16^7}$		\vspace{1pt} 	\\
$ \left. \begin{array}{ccc}
	\makebox[18pt][c]{$a + b$}		\phantom{\hspace{10pt}}		& 1110 		& \hspace{11pt} 	3	\\
	\makebox[18pt][c]{$a + \nb$}		\phantom{\hspace{10pt}}		& 1101 		& \hspace{11pt} 	3	\\
	\makebox[18pt][c]{$\na + b$}		\phantom{\hspace{10pt}}	 	& 1011 		& \hspace{11pt} 	3	\\
	\makebox[18pt][c]{$\na + \nb$}		\phantom{\hspace{10pt}}	  	& 0111 		& \hspace{11pt} 	3		
\end{array} 	\hspace{5pt} 			\right\} $	& $\frac{1}{16}$		& $\frac{216}{16^3}$		& $\frac{42048}{16^5}$	& $\frac{8087040}{16^7}$		\vspace{1pt} 	\\
$ \left. \begin{array}{ccc}
	\makebox[18pt][c]{true}			\phantom{\hspace{10pt}}		& 1111		& \hspace{11pt} 	4
\end{array}	\hspace{5pt} 			\right\} $	& $\frac{1}{16}$		& $\frac{680}{16^3}$		& $\frac{261056}{16^5}$	& $\frac{83663360}{16^7}$ \vspace{6pt} \\ 
\end{tabularx}
\caption{\small
\textbf{Probability of output functions.}
For $k = 1$ and 2 arguments, we show the output function distribution $\p(n)$ at several network depths $n$. 
Functions with the same Hamming weight $w$ have the same probability.
The probability shown for each $w$ group is that of each individual function in the group.
At $n=0$ all functions are equally likely;
as depth increases, functions with low or high $w$ become more probable.
The functions are shown algebraically and by their truth tables.
In our notation, 
$\na$ means {\sc not} $a$, 
$ab$ means $a$ {\sc and} $b$, 
$a \oplus b$ means $a$ {\sc xor} $b$, and
$a + b$ means $a$ {\sc or} $b$.
\vspace{-10pt}
}
\label{k123Distribution}
\end{table} 
\newpage \noindent {\sf\textbf{\textcolor{black}{Exact distribution}}}  \\
We use $f$ for the local Boolean functions assigned to the network nodes and $F$ for the global function computed by the entire network.
Whereas $f$ is a function of the $k$ outputs of the layer immediately below it, $F$ is a function of the $k$ terminal arguments \(a,b,\ldots\) (Fig. \ref{Architecture}).
\\ \indent
Before solving the general case, let's consider the output distribution for small $k$ and $n$.
For $k=1$ and $2$, the 4 and 16 possible output functions are shown on the left of Table~\ref{k123Distribution}, both algebraically and by their truth tables.
A truth table is a binary string of length $2^k$ specifying the function's output for every combination of its $k$ inputs, where 0 is false and 1 is true.
We group functions with the same Hamming weight $w$, that is, the number of 1s in the truth table.
On the right of Table \ref{k123Distribution} are the probabilities of the different global functions $F$ for various $n$.
These start out uniform but become highly nonuniform as $n$ increases.
\\ \indent
As a concrete example, take $k=2$ and $n=1$.
If $f_0$ is {\sc and}, $f_1^{(1)}$ is {\sc or}, and $f_1^{(2)}$ is {\sc nand}, then
$ F(a,b) \allowbreak
	= f_0\big(f_1^{(1)}(a,b),f_1^{(2)}(a,b)\big) \allowbreak
	= (a+b)(\na+\nb) \allowbreak
	= a\nb+\na b \allowbreak
	= a\oplus b$.
Repeating this process over the $16^3$ possible assignments to the three local functions gives the distribution of $F(a,b)$ shown in Table~\ref{k123Distribution}B.
\\ \indent
We now turn to the general solution for arbitrary $k$ and $n$.
We increase the depth by adding a new layer of $k$ random local functions at the bottom of the machine, whose outputs become the $k$ arguments of $F_{n-1}$.
Thus
\begin{align*}
	F_n(a,b,\ldots)
	=
	F_{n-1}\!\left(
	f_n^{(1)}(a,b,\ldots),\ldots,
	f_n^{(k)}(a,b,\ldots)
	\right).
\end{align*}
\indent
For any fixed input configuration $(a,b,\ldots)$, the outputs
$f_n^{(1)}(a,b,\ldots),\ldots,f_n^{(k)}(a,b,\ldots)$
are independent, and each is $0$ or $1$ with probability $1/2$.
These $k$ outputs therefore form one of the $2^k$ possible input configurations of $F_{n-1}$, each with probability $1/2^k$.
Since $F_{n-1}$ returns $1$ for $w(F_{n-1})$ of these configurations, the corresponding output of $F_n$ is $1$ with probability $w(F_{n-1})/2^k$.
Writing the $2^k$ entries of the truth table of $F_n$ as $\sigma_1,\ldots,\sigma_{2^k}$, the $\sigma_i$ are thus independent and identically distributed, with
\begin{align*}
	P(\sigma_i=1) &= \frac{w(F_{n-1})}{2^k},
	\qquad
	P(\sigma_i=0) = 1-\frac{w(F_{n-1})}{2^k}.
\end{align*}
\indent
For convenience, let $\ell=2^k$.
The probability of obtaining a particular $F_n$, given $F_{n-1}$, is then
\begin{align*}
	P(F_n\mid F_{n-1})
	=
	\prod_{i=1}^{\ell}
	\left(\frac{w(F_{n-1})}{\ell}\right)^{\sigma_i}
	\left(1-\frac{w(F_{n-1})}{\ell}\right)^{1-\sigma_i}.
\end{align*}
Noting that $\sum_i \sigma_i=w(F_n)$, and summing over $F_{n-1}$, we obtain
\begin{align*}
	P(\f_n) \!=\! \frac{1}{\ell^\ell} \! \sum_{\mathclap{\f_{n-1}}} \! w(\f_{n-1})^{w(\f_n)}  \!\left(\ell \!-\! w(\f_{n-1})\right)^{\ell - w(\f_n)} \!\! P(\f_{n-1}),
\end{align*}
where $0^0=1$, a common convention in combinatorics.
\\ \indent
Let $\p(n)$ be the probability distribution of the $2^\ell$ output functions $F_n$, in lexicographical order.
For example, for $k=1$, 
$\p(0) = (\nicefrac{1}{4}, 		\allowbreak \nicefrac{1}{4}, 	\allowbreak \nicefrac{1}{4}, 	\allowbreak \nicefrac{1}{4})$ and
$\p(1) = (\nicefrac{6}{4^2}, 	\allowbreak \nicefrac{2}{4^2}, 	\allowbreak \nicefrac{2}{4^2}, 	\allowbreak \nicefrac{6}{4^2})$,
with more examples in Table \ref{k123Distribution} (though not in lexicographical order).
The elements of $\p(n)$ satisfy
\begin{align*}
	p_i(n) = \frac{1}{\ell^\ell} \sum_{j=1}^{2^\ell} w_{j}^{w_i}  \left(\ell - w_{j}\right)^{\ell - w_{i}} p_j(n-1),
\end{align*}
where $w_{i} = 0,1,1,2,1,2,2,3,\ldots$ is the Hamming weight of the binary representation of $i-1$,
and $p_i(0) = 1/2^\ell$.
\\ \indent
Notice that the probability of an output at level $n$ depends on an output at level $n-1$ only through its Hamming weight.
We can therefore replace the distribution $\p$ over individual output functions by the simpler distribution $\q$ over Hamming weight.
The symmetry in $\p$ reduces the problem from $2^{2^k}$ output probabilities to $2^k+1$ Hamming-weight probabilities.
For example, 
for $k=1$, 
$\q(0) = (\nicefrac{1}{4}, \nicefrac{2}{4}, \nicefrac{1}{4})$ and
$\q(1) = (\nicefrac{6}{4^2}, \nicefrac{4}{4^2}, \nicefrac{6}{4^2})$,
with more examples in Table \ref{k123Distribution}.
The distribution of the Hamming weight satisfies
\begin{align*}
	q_i(n) = \frac{1}{\ell^\ell} \sum_{j=0}^\ell \binom{\ell}{i} j^i  \left(\ell - j\right)^{\ell - i} q_j(n-1).
\end{align*}
In other words, $\q(n) = \A \q(n-1)$,
where $\A$ is the $\ell + 1$ by $\ell + 1$ transition matrix
	\begin{align*}
		A_{i,j} 	& =	\frac{1}{\ell^\ell}		\binom{\ell}{i}	j^i (\ell-j)^{\ell-i}.
	\end{align*}
We index the rows and columns of $\A$ from $0$ to $\ell$,
and recall that $\ell=2^k$ and $0^0=1$.
This matrix representation immediately gives
\begin{align}
	\q(n) 	& =	\A^{\!n}	\q(0),
	\label{MainEquation}
\end{align}
where $q_i(0) = \binom{\ell}{i}\big/2^\ell$.
\\ \indent
To get a sense of the structure of $\A$, it helps to see some examples. 
For $k=1$,
\begin{align*}
\A & = 
	\begin{scriptsize}
	\frac{1}{2^2} 
	\begingroup
		\setlength\arraycolsep{0pt}
		\left(	
			\begin{array}{ccc}
			\binom{2}{0}	& 				& 				\vspace{1pt}	\\
						& \binom{2}{1}		& 				\vspace{1pt}	\\
						&				& \binom{2}{2} 		
			\end{array}
		\right)
	\endgroup
	\begingroup \setlength\arraycolsep{2.5pt} 
		\left(		\begin{array}{ccc}
		0^0 2^2 	& 1^0 1^2 		& 2^0 0^2 		\vspace{2pt}  \\
		0^1 2^1 	& 1^1 1^1 		& 2^1 0^1 		\vspace{2pt}  \\
		0^2 2^0 	& 1^2 1^0 		& 2^2 0^0 			
		\end{array}	\right).
	\endgroup
	\end{scriptsize}
\end{align*}
For $k=2$,
\begin{align*}
\A & = 
	\begin{scriptsize}
	\frac{1}{4^4} 
	\begingroup
		\setlength\arraycolsep{0pt}
		\left(	
			\begin{array}{ccccc}
			\binom{4}{0}	& 				& 				& 				&				\vspace{1pt}	\\
						& \binom{4}{1}		& 				& 				&				\vspace{1pt}	\\
						&				& \binom{4}{2} 		& 	 			&				\vspace{1pt}	\\
						&				& 				& \binom{4}{3} 		&				\vspace{1pt}	\\
						&				& 				& 				& \binom{4}{4} 		
			\end{array}
		\right)
	\endgroup
	\begingroup \setlength\arraycolsep{2.5pt} 
		\left(		\begin{array}{ccccc}
		0^0 4^4 	& 1^0 3^4 		& 2^0 2^4 		& 3^0 1^4		& 4^0 0^4		\vspace{2pt}  \\
		0^1 4^3 	& 1^1 3^3 		& 2^1 2^3 		& 3^1 1^3		& 4^1 0^3		\vspace{2pt}  \\
		0^2 4^2 	& 1^2 3^2 		& 2^2 2^2 		& 3^2 1^2 		& 4^2 0^2		\vspace{2pt} \\
		0^3 4^1 	& 1^3 3^1		& 2^3 2^1 		& 3^3 1^1 		& 4^3 0^1		\vspace{2pt} \\
		0^4 4^0 	& 1^4 3^0 		& 2^4 2^0 		& 3^4 1^0 		& 4^4 0^0	
		\end{array}	\right).
	\endgroup
	\end{scriptsize}
\end{align*}
\\ \indent
For most of this paper we work with $\q$, the distribution of output Hamming weights.
But we're ultimately interested in $\p$, the distribution of individual output functions.
To translate between them, 
\begin{align*}
	p_j = \textstyle q_i \big/ \binom{2^k}{i},
	\label{Translate}
\end{align*}
for all $j \in [1,2^\ell]$ such that $w(F_j) = i$.%
\\ \\ \noindent 
{\sf\textbf{\textcolor{black}{Spectral structure}}}  \\
The matrix $\A$ has $2^k + 1$ eigenvalues and eigenvectors.
The first two we can work out for free.
That's because the first and last columns of $\A$ are $(1, 0, \ldots, 0)$ and $(0, \ldots, 0, 1)$.
So these are the first two eigenvectors, with corresponding eigenvalues $\lambda_0 = \lambda_1 = 1$.
\\ \indent
More generally, the $r$th eigenvalue is
\begin{align*}
	\lambda_r = \frac{(\ell)_r}{\ell^r},
\end{align*}
where $\ell = 2^k$ and  $(\ell)_r = \ell (\ell - 1) \ldots (\ell - r + 1)$ is the falling factorial.
The spectrum follows directly from the binomial form of the columns of $\A$. 
Since $\A$ and $\A^T$ have the same eigenvalues, 
we can obtain the spectrum by applying $\A^{T}$ to the falling-factorial basis $\bigl((i)_0,(i)_1,\ldots,(i)_\ell\bigr)$.
Using the factorial-moment identity for the binomial distribution,
	\begin{equation*}
		\sum_{i=0}^{\ell} \! A_{i,j}(i)_r 	= \frac{(\ell)_r}{\ell^r}j^r
								= \frac{(\ell)_r}{\ell^r}(j)_r + \raisebox{-0.26\height}{\shortstack{lower-degree\\terms}}.
	\end{equation*}
Thus $\A^T$ is triangular in the falling-factorial basis, with diagonal entries $\lambda_r={(\ell)_r}/{\ell^r}$,
which are therefore the eigenvalues of $\A$.
\\ \indent
In the limit of large network depth, the first two eigenvectors dominate, 
so the probabilities of all non-constant outputs vanish and the output becomes a toss between true and false. 
But this simple asymptotic picture conceals the finite-depth structure of the bulk. 
Its shape is governed by the leading transient eigenvector $\mathbf{v_2}$.
Although $\v_{\mathbf 2}$ has no simple finite-$\ell$ form, its bulk profile can be determined exactly in the large-$\ell$ limit.
\\ \indent
Again let $x = i/\ell$, and write the large-$\ell$ limit of the bulk of $\mathbf{v_2}$ as
$v_{2,i} = V(x)/\ell$. 
The binomial kernel of $\A$ has zero drift and variance $x(1-x)/\ell$, so in the bulk
	\begin{align*} 
		\A = \mathbf{I} + \frac{1}{\ell} \mathbf{G} + O(\ell^{-2}), \qquad
		\mathbf{G} \phi = \frac12\frac{d^2}{dx^2} \big(x(1-x) \phi(x) \big).
	\end{align*} 
Since
$\lambda_2=1-1/\ell$, the equation $\A \v_\mathbf{2} = \lambda_2 \v_\mathbf{2}$ gives
	\begin{align*}
		\frac{1}{2}\frac{d^2}{dx^2}\big(x(1-x)V(x)\big) = -V(x).
	\end{align*}
Expanding the derivative and eliminating $V$, we can write
	\begin{align*}
		\frac{d}{dx} \big(x^2(1-x)^2V'(x)\big) =0.
	\end{align*}
So $x^2(1-x)^2V'(x) = C$, a constant.
Since by reflection symmetry $V(x) = V(1-x)$, the midpoint is stationary, 
so $V'(1/2) = 0$ and $C = 0$.
Since $x^2 (1 - x)^2 > 0$ for $x \in (0,1)$, $V' = 0$, so $V$ is flat.
\\ \indent
Now we just need to determine its value. 
Since $v_{2,0} = v_{2,\ell}=-1$ and the components of $\v_{\mathbf{2}}$ sum to zero,
we have
\[
\lim_{\ell\to\infty}
\left(
-2+\frac{\ell-1}{\ell}V
\right)
=
0,
\]
so $V = 2$.
Therefore in the limit of large $\ell$, for $0 < i/\ell < 1$,
	\begin{align*}
		v_{2,i} = 2/\ell.
	\end{align*}
Thus the leading transient mode is flat across Hamming-weight classes in the bulk.
\begin{figure}[!b]
	\includegraphics[width=1\columnwidth]{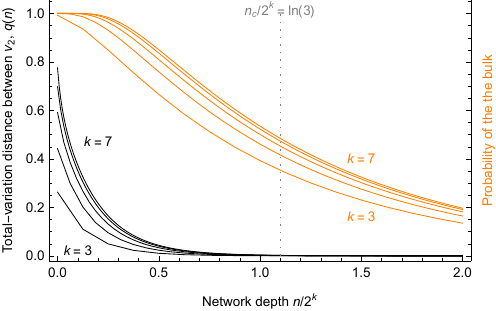}
	\caption{\small
	\textbf{Bulk relaxation and crossover.}
	The orange curves show the total bulk probability, 
	which decays as probability accumulates in the absorbing states true and false. 
	The black curves show the total-variation distance between the normalized interior of $\q(n)$ and that of $\mathbf{v_2}$. 
	The leading deviation from $\mathbf{v_2}$ decays as $\exp(-5n/2^k)$, 
	consistent with the rapidly decreasing black curves, whereas the total bulk probability decays as $\exp(-n/2^k)$.
	Thus the bulk acquires its dominant shape well before it disappears. 
	The vertical line marks the crossover depth $n_{\rm c} \simeq \ln 3\,2^k$,
	where true and false together carry half the probability.
	}
    \label{CriticalPlot}
\end{figure}
\\ \\ \noindent\!\!{\sf\textbf{Finite-depth crossover}} \\
As the network depth $n$ increases, the output distribution $\p$ changes in two competing ways. 
First, the bulk of $\p$ shifts from uniform to {\sf U}-shaped with respect to Hamming weight
(Fig. \ref{ComputerExperiment}).
Outputs with low or high Hamming weights become exponentially more likely than those with intermediate ones. 
Second, the probability of the constant functions true and false shifts from negligible ($1/2^{2^k}\!\!$ each) to dominant ($\nicefrac{1}{2}$ each),
causing the combined probability of all of the other functions to vanish.
\\ \indent
Strikingly, this competition produces a crossover at depth of order $2^k$. 
The bulk acquires its characteristic shape before its probability substantially diminishes; at greater depth, loss into true and false dominates and the distribution progressively collapses onto these two absorbing states. 
\\ \indent
To estimate the crossover depth, we retain the three slowest modes in the spectral expansion of $\q(n)$.
The first two eigenvectors describe the absorbing endpoints, 
while the leading transient eigenvector $\mathbf{v_2}$ governs the bulk.
Then
	\begin{align}
		\q(n)	& \simeq c_0 \vi \lambda_0^n + c_1 \vii \lambda_1^n + c_2 \lambda_2^n \mathbf{v_2} 	\nonumber \\
			& \simeq c_0 \vi + c_1 \vii + c_2 e^{-n/\ell} \mathbf{v_2},
			\label{HA}
	\end{align}
since $\lambda_0 = \lambda_1 = 1$ and $\lambda_2^n = (1 - \nicefrac{1}{\ell})^n \simeq e^{-n/\ell}$.
\\ \indent
Now let's work out the projection coefficients.
At $n=0$, all $2^\ell$ output functions are equally likely, each with probability $1/2^{\ell}$. 
Since there are $\binom{\ell}{i}$ functions of Hamming weight $i$, the initial Hamming-weight distribution is
$\q_i(0)=\binom{\ell}{i}/2^\ell$.
Projecting this initial condition onto the eigenvectors gives $c_0=c_1=\nicefrac12$. 
The coefficient $c_2$ can be calculated in the large-$\ell$ limit as follows.
Consider the left eigenvector $\mathbf{u_2}$ of $\A$.
Unlike the right eigenvector $\mathbf{v_2}$, we can readily calculate it: 
$u_{2,i} = i (\ell - i)$. 
Then
\begin{equation*}
c_2 = \frac{\mathbf{u_2} \q(0)}{\mathbf{u_2} \mathbf{v_2}} 
	\simeq \frac{\ell (\ell - 1)/4}{\ell (\ell + 1)/3} = \frac{3 (\ell - 1)}{4 (\ell + 1)} \simeq \frac{3}{4}.
\end{equation*}
Here the numerator $\ell (\ell - 1)/4$ is exact, 
whereas for the denominator we use the flat finite-$\ell$ approximation $v_{2,i} \simeq 2/(\ell-1)$, which preserves the exact interior sum.
Then eq. (\ref{HA}) becomes
\begin{align*}
	\q(n)		& \textstyle \simeq (\frac{1}{2} , 0, \ldots, 0, \frac{1}{2} ) \textstyle + \nicefrac{3}{4} \, e^{-n/\ell} (-1, \frac{2}{\ell - 1}, \ldots, \frac{2}{\ell - 1}, -1).
\end{align*}
\indent 
The crossover depth occurs when the two endpoints and the bulk have equal probability, giving 
	\begin{align*}
		n_{\rm c} \simeq \ln(3) 2^k.
	\end{align*}
This is marked in Fig. \ref{CriticalPlot}.
\\ \indent
Importantly, the normalized bulk of $\q$ approaches $\mathbf{v_2}$ much faster than its total probability decays.
Since the transition matrix and the binomial initial condition are symmetric under \(i\leftrightarrow \ell-i\), 
the projections onto the antisymmetric eigenvectors vanish, so \(c_3=c_5=\cdots=0\). 
The leading correction is therefore $\mathbf{v_4}$, whose amplitude relative to $\mathbf{v_2}$ decays as
	\begin{equation*}
		\left ({\lambda_4}/{\lambda_2}\right)^n=\big((1-{2}/{\ell})(1-{3}/{\ell})\big)^n\simeq e^{-5n/\ell}.
	\end{equation*}
By contrast, the total bulk probability decays as \(e^{-n/\ell}\). 
Thus the bulk acquires its characteristic shape well before it disappears, 
creating a distinct finite-depth regime before collapse onto true and false.
\\ \\ \noindent {\sf\textbf{Comparison with experiments}}
\\ To confirm our theoretical prediction of the output distribution,
we conducted extensive experiments for various values of the number of arguments $k$ and network depth $n$.
We did complete enumeration when possible, and otherwise sampled from the ensemble of configurations.
In all cases, our computer experiments match our theory.
\\ \indent 
For $k = 2$, there are $16^3, 16^5, 16^7$ and  $16^9$ configurations for depths $n = 1, 2, 3$ and 4.
We enumerated all of these configurations and, for each, determined the network's output function.
We plot the probability $\q_i(n)$ of obtaining an output with Hamming weight $i$ in Fig. \ref{ComputerExperiment}A (points).
This exactly matches our prediction given by eq. (\ref{MainEquation}):
the solid line shows the probabilities of the bulk and 
the dashed line the outputs true ($w = 4$) and false ($w = 0$).
As $n$ increases, the likelihood of true and false approach $\nicefrac{1}{2}$ and the likelihoods of all other outputs vanish.
\\ \indent
For $k = 3$, there are $256^4$, $256^7$ and $256^{10}$ configurations for $n = 1, 2$ and 3.
For $n = 1$, we enumerated all of the configurations.
For $n = 2$ and $3$, this is computationally infeasible, so we sampled them instead.
We randomly assigned one of the 256 functions to each of the nodes and then determined the network output, repeating this two million times.
These are plotted in Fig. \ref{ComputerExperiment}B (points).
We calculated error bars, but these are negligible.
Our theory agrees with the computer experiments exactly for $n = 1$ and to within statistical significance for $n = 2$ and 3.
\\ \indent
For $k = 4$, there are $65536^5$ and $65536^9$ configurations for $n = 1$ and 2.
These are too many to enumerate, so we sampled as above.
These are plotted in Fig. \ref{ComputerExperiment}C. 
Once again, our experiments confirm our theory.
\\ \\ \noindent {\sf\textbf{Discussion}}
\\ \noindent 
We have derived the exact output distribution of a deep-layered Boolean machine at finite width and depth.
Increasing depth does not simply drive the output distribution toward its trivial infinite-depth limit.
Rather, repeated composition first creates a strongly structured distribution over the bulk (everything but true and false).
Once the leading transient mode dominates,
	\begin{equation*}
		P(F) \simeq \frac{3e^{-n/2^k}}{2^{k+1}\binom{2^k}{w(F)}}.
	\end{equation*}
At greater depth, the distribution ultimately collapses onto a coin toss between true and false.
\\ \indent 
The separation of relaxation rates creates distinct finite-depth regimes.
At first, adding layers reshapes the bulk distribution, with little probability reaching true and false; 
later, the bulk has largely acquired its characteristic form while probability increasingly drains into true and false. Mozeika, Li and Saad observed that ``random deep ReLU networks compute only constant Boolean functions in the infinite-depth limit'' \cite{Mozeika2020}. 
Our finite-depth solution agrees with this asymptotic result, 
but reveals what it conceals: 
an extended transient regime in which the bulk distribution develops its strongest structure. 
\\ \indent 
Our model is tractable because a vast microscopic state space admits an exact macroscopic description.
Although there are $2^{2^k}$ possible output functions, symmetry reduces the dynamics to $2^k+1$ Hamming-weight classes.
The transpose of the resulting transition matrix $\A$ is the conventional neutral Wright-Fisher matrix. 
Its action on polynomial observables preserves degree, 
making the eigenvalues and left eigenvectors of $\A$ readily accessible.
The finite-depth output distribution, however, evolves under $\A$, 
not $\A^{\rm T}$, and therefore depends on the much harder right eigenvectors of $\A$.
\\ \indent 
Fig. \ref{CriticalPlot} suggests that the finite-depth dynamics may become universal at large $k$. 
When depth is rescaled by $2^k$, the curves for different $k$ appear to approach a limiting form.
We conjecture that, as $k$ becomes large, the dynamics is governed by the rescaled depth $n/2^k$. 
Such a limit would replace the family of finite-$k$ machines by a single continuum description, 
providing a thermodynamic picture of deep composition.
\\ \indent 
Overparameterized neural networks have vastly more microscopic configurations than are needed to fit the training data, yet generalize well to unseen data.
A potential resolution to this puzzle is that sampling these configurations induces a highly nonuniform distribution over global functions.
Recent work by Mingard et al. \cite{Mingard2025} found empirical evidence that the function prior induced by deep neural networks helps explain their ability to generalize.
Our model isolates repeated composition as one source of such a prior.
The effective number of functions available to learning can be far smaller than the nominal size of function space.
\\ \indent 
Our exactly solvable model provides a tractable setting for understanding simplicity bias in input-output maps. 
In such maps, many microscopic descriptions generate a smaller space of macroscopic outcomes. 
Input-output maps occur across biology, physics and technology, 
and mounting empirical evidence suggests that they strongly favor simple outputs
\cite{Dingle2018, Johnston2022, Mingard2025}.
Because the output probabilities of our machine can be calculated analytically, 
it offers a rare setting in which this bias can be derived rather than just observed. 
We investigate the simplicity bias of deep-layered machines elsewhere \cite{Fink2026}.

\end{document}